\documentclass[12pt]{article}
\usepackage{amsfonts}
\usepackage{times}
\def\Ib{{\mathbb I}\,}

\def\eopp{{ \vrule height7pt width7pt depth0pt} }
\def\Prf{\noindent {\em Proof:}\, }

\def\Ib{{\mathbb{I}}\,}

\def\Ah{\hat{A}\,}
\def\Atl{\tilde{A}\,}
\def\Btl{\tilde{B}\,}
\def\Ctl{\tilde{C}\,}
\def\Ch{\hat{C}\,}

\def\Otl{\tilde{O}\,}

\def\Qtl{\tilde{Q}\,}
\def\Ctl{\tilde{C}\,}

\def\Vtl{\tilde{V}\,}

\def\Qh{\hat{Q}\,}

\def\Qctl{\tilde{\cal Q}\,}
\def\eopp{{ \vrule height7pt width7pt depth0pt} }
\def\Prf{\noindent {\em Proof:}\, }

\def\BibTeX{{\rm B\kern-.05em{\sc i\kern-.025em b}\kern-.08em
    T\kern-.1667em\lower.7ex\hbox{E}\kern-.125emX}}

\makeatletter
\@addtoreset{equation}{section}
\makeatother
\newtheorem{thm}{Theorem}[section]
\newtheorem{defn}[thm]{Definition} 
\newtheorem{lem}[thm]{Lemma} 
\newtheorem{cor}[thm]{Corollary} 
\newcommand{\BEQ}{\begin{equation}}
\newcommand{\EEQ}{\end{equation}}
\newcommand{\NEQ}{\end{equation}}
\begin{document}
\bibliographystyle{plain}
\title{ORTHONORMAL REPRESENTATIONS 
FOR OUTPUT SYSTEM PAIRS}

\author{
Kurt S.~Riedel${^\dagger}$ and Andrew P.~Mullhaupt
\thanks{
S.A.C. Capital Management, LLC,
540 Madison Ave, New York, NY 10022, \ \
$\dagger$  Millennium Partners, 666 Fifth Ave, New York, 10103-0899}
}

\date{2000}

\maketitle

\begin{abstract}
A new class of canonical forms is given proposed in which
$(A, C)$ is in Hessenberg observer or Schur form and
output normal: $\Ib - A^*A =C^*C$.
Here, $C$ is the $d \times n$ measurement matrix and $A$ is the advance matrix.
The $(C, A)$ stack is expressed as the product of $n$
orthogonal matrices, each of which depends on $d$ parameters.
State updates require only ${\cal O}(nd)$ operations and
derivatives of the system with respect to the parameters
are fast and convenient to compute.
Restrictions are given such that these models are generically identifiable.
Since the observability Grammian is the identity matrix, system identification
 is better conditioned than other classes of models with fast updates.
\end{abstract}

{\bf AMS Classifications:} 93B10, 93B11 93B30

{\bf Key words.} System representations, balanced representations,
Hessenberg observer form, Schur representations,
Stein's equation, discrete Lyapunov equation

\newpage
\section{INTRODUCTION} \label{I}

Canonical forms are important in system identification,
where an unique representation is desired to avoid identifiability
problems \cite{LS,Ober}.
We consider the system $(A, B, C)$, where $A$ is a real $n\times n$ matrix,
 $B$  is a real $n\times m$ matrix, and
$C$ is $d\times n$ real matrix with $n \ge d$.
Our systems are output normalized:
\BEQ \label{ONeq}
{ A}^* { A} =  { \Ib_n} - { C}^* {C} \ \ \ .
\NEQ
In the next section, we describe the advantages of using an input normal
or an output normal (ON) representation.

We consider real stable output pairs
and show that a real output pair representation always
exists where $(A,C)$ is simultaneously output normal and in Hessenberg
observer form or real Schur form.
We give explicit parameterizations of the $(C,A)$
stack as a product of orthogonal matrices of the form:
\BEQ \label{GenForm}
\left( \begin{array}{c} C \\ A \end{array} \right) =
\left[ \prod_{i=1}^{nd}
G_{j(i),k(i)}(\theta_{i})\right]_{1:(n+d),1:n}
\NEQ
and related variants. Here $G_{j,k}$ is a Given's rotation in ${\cal R}^{n+d}$
as defined at the end of this section. Our representations include the
banded orthogonal filters of \cite{MRbook} as a special case
(under the duality map $A \rightarrow A^*$, $C \rightarrow B$).

These orthogonal product representations are parameterized by the
minimal number of free parameters
and have no coordinate singularities.
Our representation allow fast state updates in ${\cal O}(nd)$ operations and
derivatives of the system with respect to the parameters
are fast and convenient to compute. 

Our results consider only the output pair, $(A,C)$,
and are independent of $B$.
Thus $B$ and $D$ may be treated as linear parameters
in system identification or system synthesis (in contrast to (\ref{Lossless}))
and chosen separately from the parameters of $A$ and $C$. In particular,
the elements of $B$ may be estimated with pseudo-linear regression.
Corresponding controller representations
exist for input pairs, $(A,B)$.

 Any stable observable output pair may be transformed into
one of our representations by the following three step process.
First, we transform the output pair $(A,C)$ to output normal form
using the Cholesky factor of the solution of (\ref{DSteinEq}).
Second, we orthogonally transform the output normal pair $(A,C)$
to any of the three major output forms: Schur form, Hessenberg observer form,
and observer triangular system form as defined in Section \ref{DefSect}.
Finally, we perform a series of Givens rotations to show
that the transformed system must be of the form given by (\ref{GenForm}).

The final representations are given in
Theorems \ref{OTSONRepthm}, \ref{HOONRepthm}.
For statistical estimation and numerical implementations,
it is highly desirable to eliminate redundancy
in the parameterization when possible.
We address redundancy in two ways. First, we categorize
when two distinct ON pairs in Hessenberg observer form are equivalent.
Second, we impose constraints on
the parameters in (\ref{GenForm}) to eliminate redundant parameterizations
of the same ON pair \underline{generically}.
We repeat this analysis for Schur form and for observer triangular
system form.


In Section \ref{CondSect}, we give a brief overview of the advantages of
input normal and output normal form.
In the Section \ref{DefSect}, we give the basic definitions and
show every output pair
is similar to an output pair in Schur ON form, to a Hessenberg ON pair
and to an ON pair in triangular system form .
In Section \ref{UniqSect}, 
we show that after standardization
these Hessenberg ON systems uniquely parameterize transfer functions
for {\em generic} systems. When $A$ is reducible, we find an orthogonal
transformation that preserve the output normal property.
To construct the orthogonal product representations of the
$(C,A)$ stack, we need families of orthogonal matrices such as
the set of Householder matrices.
In Section \ref{MOPSect}, we  give a general definition of
orthogonal reduction families that includes Householder and Givens representation.
In Section \ref{OTSONRepSect} and Section \ref{HOONRepSect},
we give explicit orthogonal product representations of Hessenberg
output normal pairs. 

{\em Notation:}
The $n \times n$ identity matrix is $\Ib_n$ and
$e_k$ is the unit vector in the $k$th coordinate.
By $A_{i:j,k:m}$, we denote the $(j-i+1)\times (m-k+1)$ subblock
of $A$ from row $i$ to row $j$ and from column $k$ to column $m$.
We abbreviate $A_{i:j,1:n}$ by $A_{i:j,:}$.
The matrix $A$ has upper bandwidth $d$ if $A_{i,j}=0$ when $j> i+d$.
A $k \times m$ matrix of zeros  is denoted by $0_{k,m}$.
The direct sum of matrices is denoted by $\oplus$.
We denote the matrix transpose of $A$ by $A^*$ with no complex conjugation
since we are interested in the real system case.

We denote
the Given's rotation in the $i$th and $j$th coordinate by $G_{ij}$ i.e.
$g_{i,i}=g_{j,j}= \cos(\theta)$, $g_{i,j}= -g_{j,i}= \sin(\theta)$ and
$g_{k,m} = \delta_{k,m}$ otherwise, where $g_{k,m}$ are the elements of
$G_{ij}$.
The symbol $E$
denotes a signature matrix: $E_{i,j}^2 = \delta_{i,j}$.

Two systems $(A, B, C)$ and  $(\Atl, \Btl, \Ctl)$ are similar
(equivalent) when
$\tilde{ A} \equiv { T^{-1}AT}$, $\tilde{ C}\equiv { CT}$
and  $\tilde{ B}\equiv { T^{-1}B}$
for some invertible $T$. They  are orthogonal equivalent if
$T$ is a real orthogonal matrix.

\section{REPRESENTATIONS AND CONDITION NUMBERS} \label{CondSect}

The goal of this paper is to propose system representations that are both
well conditioned for system identification and are fast and convenient for
numerical computation. We briefly discuss these issues in the context of
existing alternative  system representations. For more complete analysis
of conditioning in system identification, we refer the reader to \cite{MR4}.

Let $(A,B,C)$ be stable, observable and controllable. We define
the observability Grammian, $P_{A^*,C^*}$ and the controllability Grammian,
$P_{A,B}$ by
\BEQ \label{SteinEq}
P_{A,B} -AP_{A,B}A^*=BB^*
\NEQ
\BEQ \label{DSteinEq} \label{DualStein}
P_{A^*,C^*} -A^*P_{A^*,C^*}A=C^*C \ .
\NEQ

A popular class of system representations is balanced systems
\cite{Moore,Ober,Ober2},
where both the observability Grammian and the controllability Grammian are
simultaneously diagonal: $P_{A,B}= P_{A^*,C^*} = \Sigma_{A,B,C}$.
Balanced representations have many desirable theoretical properties.
However, existing parameterizations of balanced models require
 $O(n^2)$ operations to update the state space system.

An alternative to balanced models is output
normal (ON) representations \cite{MR1,MR2}, where
the observability Grammian is required to be the identity matrix,
but no structure on the controllability Grammian.

\begin{defn} \label{DefON}
An output pair, $({A},{ C})$, is
output normal  (ON) if and only if  (\ref{ONeq}) holds.
An input pair, $({A},B)$, is
input normal  (IN) if and only if
\BEQ \label{INeq}
{ A} { A}^* =  { \Ib_n} - { B} {B}^* \ \ \ .
\EEQ
\end{defn}

If $A$ is stable, definition \ref{DefON} is equivalent to $P_{A^*,C^*}=\Ib_n$
for output normal and $P_{A,B}=\Ib_n$ for input normal.
In \cite{Ober}, Ober shows that stability plus a positive definite
solution to the dual Stein equation, (\ref{DualStein}), implies
that the output pair is observable.
By Theorem 2.1 of \cite{AM},
if the observability Grammian is positive definite and $(A,C)$ is observable,
then the output pair is stable.
Thus for ON pairs, stability is equivalent to observability.

ON pairs are not required to be stable or observable.
(From (\ref{ONeq}), $A$ must be at least marginally stable.)
In \cite{Moore},  `output normal'' has a more restrictive
definition of (\ref{ONeq}) and the additional requirement that the
controllability Grammian be diagonal. We do not impose any such condition
on the controllability Grammian.
In  \cite{MR1}, we called condition (\ref{ONeq}) `output balanced'',
whereas now we call (\ref{ONeq}) `output normal.''
We choose this language so that `normal'' denotes restrictions on only
one Grammian while `balanced'' denotes simultaneous restrictions on
both Grammians.


A measure of ill-conditioning in system identification is the
condition number of $P_{A,B}$, $\kappa(P_{A,B})\equiv$ largest
singular value of   $P_{A,B}$ divided by the smallest.
In \cite{MR4}, we show that solving the Stein equation,
$P_{A,B}$ 
is exponentially ill-conditioned in $n/m$ for large classes of $(A,B)$ pairs;
i.e.\ ${\kappa} (P_{A,B}) \sim \exp( \alpha n/m)$ for some $\alpha$.
To avoid the possibility of ill-conditioning, we prefer to consider
representations where either the observability or the controllability
Grammian is the identity.

Let $\Sigma_{A,B,C}$ be the Grammian of the balanced system
equivalent to $(A,B,C)$.
In \cite{Laubetal}, it is shown that
\BEQ
{\kappa \left( \Sigma_{A,B,C}^{}\right)^2 }\le
\kappa _{}\left( P_{A,B}\right) \kappa\left( P_{A^*,C^*}\right) \ ,
\NEQ
where equality holds for balanced systems, input normal systems and
output normal systems.
For output balanced systems, the ill-conditioning is entirely in
the controllability Grammian:
$ \kappa _{}\left( P_{A,B}\right) = \kappa \left( \Sigma_{A,B,C}^{}\right)^2$.
We interpret ${\kappa \left( \Sigma_{A,B,C}^{}\right)^2 }$ as the
intrinsic conditioning of a linear time invariant (LTI) system
and
$\kappa _{}\left( P_{A,B}\right) \kappa\left( P_{A^*,C^*}\right)
 / {\kappa \left( \Sigma_{A,B,C}^{}\right)^2 }$ as a measure of
the excess ill-conditioning of a system representation.

Our representations resemble those based on embedded lossless
systems \cite{Desai,Vaid,VeenVib}:
\BEQ \label{Lossless}
\left( \begin{array}{cc} A  &B\\ C &D\end{array} \right) =
P_1 \prod_{i=1}^{f_1}G_{k(i),m(i)}(\theta_{i})
P_2 \prod_{j=f_1+1}^{f_2}G_{k(j),m(j)}(\theta_{j}) P_3 \ ,
\NEQ
where $f_2=$ number of free parameters, $P_1$ and $P_3$ are
projections onto coordinate directions and $P_2$ is a prescribed permutation.
In \cite{VeenVib}, the full system is first embedded in a lossless system
(just as  we transform the output pair $(A,C)$ to output normal form).
Next, these authors transform $(A,B)$ to Hessenberg controller form
(analogous to our transformation to Hessenberg observer form).
We conjecture that there are analogous versions of (\ref{Lossless}),
where $A$ is in Schur form or $(A,B)$ is in controller triangular system form.
Finally, the authors perform a series of Givens rotations to show
that the transformed system must be of the form given by (\ref{Lossless}).
Our corresponding representations are given in
Theorems \ref{OTSONRepthm}, \ref{HOONRepthm}. 

The main advantage of (\ref{GenForm}) over (\ref{Lossless}) is that
the observability Grammian of ON models does not inflate
the product condition number:
 $\kappa _{}\left( P_{A,B}\right) \kappa\left( P_{A^*,C^*}\right)$.
A second advantage is that $B$ and $D$ may be treated as linear parameters
in system identification or system synthesis, whereas  (\ref{Lossless})
couples the parameterization of $B$ and $D$ to that of $A$ and $C$ in
a nonlinear fashion.
For these reasons, we recommend output normal representations over embedded
lossless representations.


Another difference between our treatment and the analyses of
\cite{Desai,Vaid,VeenVib} is that we try to impose constraints on
the parameters to eliminate redundant representations whenever possible
and to categorize when redundant representations can occur.
If one is satisfied with having representations with a finite multiplicity
of equivalent systems (at least generically), this last step may
be too detailed. For numerical implementations, we believe
that it is highly desirable to eliminate as much of the redundancy
in representation as is possible.

Our representations include the banded orthogonal filters of \cite{MRbook}
as a special case. Our analysis imposes additional constraints on representations
of \cite{MRbook} to remove multiple representions of the same transfer function
generically.

\section{DEFINITIONS AND EXISTENCE}
\label{DefSect}

We now define observer triangular system form, Schur form
and Hessenberg observer form and
show that any stable observable output pair is equivalent
to an output normal pair in any of these three forms.
We denote the $(n+d) \times n$ matrix stack of $C$ and $A$ by $Q$:
\begin{equation} \label{Qdef}
Q \equiv \left( \begin{array}{c} C \\ A \end{array} \right) \ .
\end{equation}

\begin{defn} \label{DefHess}
The  output pair is in observer triangular system (OTS) form
if the $(C, A)$ stack, $Q$, satisfies
$Q_{i,j}=0$ for $j>i$. 
The output pair is unreduced if  $Q_{i,i}\ne 0$ and is reducible
if $Q_{k,k}=0$ for some $k$.
The output pair is
in standard OTS form if  $Q_{i,i}\ge 0$
and in strict OTS form if  $Q_{i,i}> 0$.
\end{defn}

Thus strict is equivalent to unreduced and standard.
The real Schur representation is defined and described
in \cite{Demmel,Edelman,GVL}. The diagonal subblocks of
$A$ may be placed in an arbitrary order.
To ensure identifiability of our model, we must specify a particular
standardization of the diagonal of the Schur form of  $A$.
Our choice, `ordered qd'' Schur form, is defined in Appendix A.

The OTS form includes the banded orthogonal filters of \cite{MRbook}
as a special case
under the duality map $A \rightarrow A^*$, $C \rightarrow B$.
Our results correspond to a detailed analysis of the generic identifiability
of the representations of \cite{MRbook}.

Hessenberg observer (HO) form is a canonical form
where $A$ is Hessenberg. We impose the additional restriction
that $C_{1,1} \ge 0$, $C_{1,j}=0$ for $j>1$. 

\begin{defn} \label{DefHOess}
The output pair is in  Hessenberg observer (HO) form
if $A$ is a Hessenberg matrix
and $C_{1,j} =0$ for $j>1$.
A HO output pair is nondegenerate if $|C_{1,1}|<1$.
A HO output pair is unreduced
if $A_{i+1,i} \ne 0 $ for $1 \le i <n$ and $C_{1,1} \ne 0$.
A HO output pair is standard if 
$A_{i+1,i} \ge 0 $ for $1 \le i <n$, $0 \le C_{1,1}<1 $.
A HO output pair is strict if it is unreduced and standard.
A HO output pair is in partial ordered Schur {\em qd} block form
if $A_{i+1,i} = 0 $ implies $A_{(i+1):n,(i+1):n}$ is in
ordered Schur {\em qd} block form.
\end{defn}

Both Hessenberg observer output pairs and
observer triangular system output pairs always can be transformed
to a standard output pair using a signature matrix, $E$:
${ A} \rightarrow { EAE}^{-1},  C \rightarrow { CE^{-1}}$.
Generically, HO output pairs are unreduced and thereby unaffected
by the requirement of partial Schur order.
For both OTS form and HO form,
the $B$ matrix is unspecified. Dual definitions for
controller forms reverse the roles of $(A,B)$ and $(A^*, C)$.
In our definitions, for $d=1$, a OTS output pair has
a lower Hessenberg matrix.


An important result in systems representation theory is

\begin{thm}\label{HOCon} \cite{LaubL,VDV}
Any observable output pair is {\rm orthogonally} equivalent to
a system in  real Schur form, to a system in observer triangular system form
and to a system in Hessenberg observer form. The Hessenberg observer form
can be chosen in partial ordered Schur {\em qd} block form.
\end{thm}

The standard proof of Theorem \ref{HOCon} begins by transforming
$C$ to its desired form and then defines Householder or Givens
rotations which zero out particular elements in $A$ in
successive rows or columns \cite{GVL}.

\begin{defn} \label{DefHON}
An output pair,  $({ A},{ C})$, is observer triangular system output normal
(OTSON) if it is in observer triangular system form and output normal.
The output pair is Hessenberg observer  output normal (HOON)
if it is in Hessenberg observer form and output normal.
The output pair is in Schur ON form if it is output normal
and $A$ is in real Schur form.
\end{defn}




\begin{thm}
\label{MONthm}
Every  stable, real observable output pair
$(A, C)$, is similar
to a real OTSON  pair, to a real HOON pair, and
to an ordered real Schur output pair with {\em qd} diagonal subblocks.
\end{thm}

\Prf
The unique solution, $P_{A^*,C^*}$, of dual Stein equation, (\ref{DSteinEq}),
is strictly positive definite. 
Let $L$ be the  unique Cholesky lower triangular factor of ${ P}$
with positive diagonal entries:
${ P} ={ LL}^{*}$.
We set  $T=L^{-*}$.
Let $U$ be orthogonal transformation that takes $(T^{-1}AT, CT)$
to the desired form (Schur, OTS or HO)
as described in \cite{LaubL,VDV}. Then
$UT$ is the desired transformation.
\eopp

This result applies to any output pair with a positive define solution to
the dual Stein equation, (\ref{DualStein}). Observability and
stability of $(A,C)$ are sufficient but not necessary conditions for
a positive definite solution.

Degenerate HOON pairs correspond to the direct sum of an identity
matrix and a nondegenerate HOON system:

\begin{lem}
Every  stable, real observable output pair
$(A, C)$, is similar to a real HOON pair with a $Q$ stack
of the form
$\Qtl = \Ib_m  \oplus \Qh$
for some $m \le d$, where $\Qh$ is a nondegenerate HOON stack.
\end{lem}

Thus we consider only HOON systems that are nondegenerate.
Note that degenerate Hessenberg controller forms are excluded from
\cite{VeenVib} by their assumptions.
If the HO pair is reducible, then it may be further
simplified using orthogonal transformations as
described in Theorem \ref{pHoQ}. 

\section{UNIQUENESS OF STRICT HOON AND OTSON REPRESENTATIONS}
\label{UniqOvSect} \label{HOONUniqSect}
\label{UniqSect}

There are two main ways in which one of our system representations can fail to
parameterize linear time invariant systems in a bijective fashion.
First, there may be a multiplicity of equivalent HOON systems
(or OTSON systems or Schur OB systems). Second, Givens product representation
such as (\ref{GenForm}) may have multiple (or no) parameterizations
of the same output pair.

For Schur OB pairs, the basic result is straightforward. If $A$ has distinct
eigenvalues and they are ordered in an unique fashion, then there is a
 parameterization that is globally bijective.

Each strict OTSON (HOON) pair generates $2^n$ distinct
but equivalent OTSON (HOON) pairs using different signature matrices.
If the OTS pair or the HO pair is reducible, then it may be further
simplified using orthogonal transformations. For the HO pair, these reductions are
described in Theorems \ref{pHoQ}. 
The representations of degenerate HO pairs reduce to a direct sum of a
`nondegenerate'' HOON system and a trivial system and thus we consider
only nondegenerate HOON pairs.

For OTSON pairs and nondegenerate HOON pairs, we find that the set of strict
output pairs has a bijective representation in an easy to parameterize subset
of  Givens product representations.
Our precise OTSON result is

\begin{thm} \label{OTSUniq}
If $(A, C)$ is a strict OTSON pair, then there are no other
equivalent strict OTSON pairs.
\end{thm}

This result and a generalization that reducible OTSON pairs is proven in \cite{MR5}.
For HOON representations, our uniqueness results are based on the following
lemma that generalizes the Implicit Q theorem \cite{Demmel,GVL}
to HOON pairs:

\begin{lem} \label{HoQ}
Let $(A, C)$ and $(\Atl, \Ctl)$ be
equivalent standard nondegenerate HOON pairs
(${\Atl} \equiv { T^{-1}AT}, { \Ctl}\equiv { C T}$).
Let $A_{k+1, k}= 0$, $C_{1,1}>0 $ and  $A_{j+1,j}>0$ for $j<k$,
then $T= \Ib_{k} \oplus U_{n-k}$, where $ U_{n-k}$ is an
$(n-k) \times (n-k)$ orthogonal matrix. Furthermore, $k>1$
and $\Atl_{k+1, k}= 0$.
\end{lem}

Since $C_{1,j}= \Ctl_{1,j}=0 $, $j>1$, $T_{j,1}= \delta_{j,1}$.
The result follows from  the Implicit Q theorem \cite{Demmel,GVL}.
\eopp

\begin{cor} \label{HoUniq}
If $(A, C)$ is a strict nondegenerate HOON pair, then there are no
other equivalent strict HOON pairs.
\end{cor}

For reducible HOON pairs, we place the lower part of $A$ in
 ordered Schur {\em qd} block form to remove redundant
representations:

\begin{thm} \label{pHoQ}
Let $(A, C)$  be a nondegenerate HOON pair with $A_{k+1,k}=0$ with $A_{j+1, j}\ne 0$ for $j<k$ and
define $A^{(2,2)}\equiv A_{(k+1):n,(k+1):n} $.
There exists an equivalent HOON pair $(\Atl, \Ctl)$,
(${\Atl} \equiv { U^{*}AU}, { \Ctl}\equiv { C U}$), where
$\Atl$ is in partial ordered Schur {\em qd} block form.
If $\Atl^{2,2}$ has distinct eigenvalues, $\Atl$ is uniquely defined.
\end{thm}



\Prf
By results cited in Appendix A, there exists an
 $(n-k) \times (n-k)$ orthogonal transformation, $U$, such that
 $U^*A^{(2,2)}U$ is in partial ordered Schur {\em qd} block form.
From Lemma \ref{HoQ},  $V= \Ib_{k} \oplus U_{n-k}$ is the desired
transformation and it is unique when $A^{(2,2)}$ has distinct eigenvalues.
\eopp


\section{ORTHOGONAL FAMILIES} \label{MOPSect}

We rewrite (\ref{ONeq}) as $Q^* Q = \Ib_{n}$,
where $Q$ is the $(n+d) \times n$ matrix stack.
Thus $Q$ is the first $n$ columns of the product of
$n$ orthogonal $(n+d)\times (n+d)$ matrices.
We parameterize each of the $n$ matrices
with $d$ parameters for a Householder transformation or $d$ Givens rotations.
We denote the group of orthogonal $m \times m$ matrices by $O(m)$.

Our basic building block is a $d$ dimensional parameterization
$ \{ \Qtl(\theta) \}$ of these orthogonal reduction transformations.
Here $\theta$ is the $d$-dimensional parameter vector.

\begin{defn} \label{ORPdef}
An orthogonal reduction parameterization (ORP)  of $O(m)$
to $e_k$ is a $m-1$ dimensional family
${\cal Q} = \{ \Qtl(\theta) \}$
of $m\times m$ orthogonal matrices
such that for every $m$ vector, $h$, there exists an {\em unique}
        $\theta(h)$
such that $\Qtl(\theta)^* h = \|h \| e_k$.
A family of orthogonal matrices is an {\em unsigned}
orthogonal reduction parameterization (ORP)  of $O(m)$
to $e_k$ if for every $m$ vector, $h \ne 0$, there exists an {\em unique}
 $\theta(h)$ such that $\Qtl(\theta)^* h$ is in the $e_k$ direction.
\end{defn}

Unsigned ORPs require that $\theta(-h) = \theta(h)$ while
standard ORPs require that $Q(\theta(-h)) \ne Q(\theta(h))$.
The $K$th column of $\Qtl(\theta)$ is equal to $\pm h / \|h \|$.
Thus $\Qtl(\theta)$ may be
determined by the $k$th column of $\Qtl(\theta)$.
For OTSON representations, we will use ORPs of $O(d+1)$ to $e_1$.
For HOON representations, we will use ORPs of $O(d+1)$ to $e_{d+1}$
and ORPs of $O(d)$ to $e_d$.

The traditional vector reduction families
are the set of Householder transformations
and families of Given's transformations. For ORPs from $O(d+1)$
to $e_1$, the two traditional Given's ORPs are
\BEQ
{\Qctl}_1 = \{ \Qtl(\theta)=
        G_{1,d+1}(\theta_d)G_{1,d}(\theta_{d-1}) \cdots G_{1,2}( \theta_1)
\} \ ,
\NEQ
\BEQ \label{ORP2}
{\Qctl}_2 = \{ \Qtl(\theta)=
G_{d,d+1}(\theta_d) G_{d-1,d}(\theta_{d-1})\cdots G_{2,3}(\theta_2)G_{1,2}(\theta_1)
 \} \ .
\NEQ
For both $\Qctl_1$ and $\Qctl_2$, we restrict the Givens angles:
 $-\pi /2 < \theta_i \le  \pi /2$ for $1<i \le d$
and $-\pi  < \theta_1 \le  \pi $. The rightmost Givens rotation
has twice the angular domain since it is used to make $e_1^*Q^*(\theta)h$ positive.

Let $\Qtl(\theta)$ be a ORP from $O(d+1)$ to $e_1$ with
the block representation:
\BEQ \label{QtlStruct}
\Qtl(\theta) =
\left( \begin{array}{cc}
\mu& y^* \\
x & \Otl
\end{array}\right) \ ,
\NEQ
where $\mu$ is a scalar and $x$ and $y$ are $d$-vectors.
The orthogonality of $\Qtl$ implies
$\mu^2 + \|x \|^2= \mu^2 + \|x \|^2 =1$,
$\mu x = - \Otl y$, $\mu y = - \Otl^* x$ and
$\Ib_d = \Otl \Otl^* + x x^* = \Otl [\Ib_d + y y^* / \mu^2 ]   \Otl^*$.
Thus $\Otl$ is invertible if $\mu \ne 0$.

We embed $\Qtl(\theta)$ in the space of $(n+d)\times (n+d)$ matrices.
\BEQ \label{QkStruct}
Q^{(k)}(\theta) = \Ib_{k-1} \oplus
\left(
\begin{array}{ccc}
 \mu_k &0_{1,n-k} &  y_k^* \\
 0_{n-k,1} & \Ib_{n-k} & 0_{n-k,d} \\
 x_k & 0_{d,n-k} &\tilde{O}_k
\end{array}\right) \ ,
\NEQ
where $ \mu_k$, $x_k$, $y_k$ and $\Otl_k$ are subblocks of
(\ref{QtlStruct}).
For the Givens rotations of class ${\cal Q}_1$, we have
\BEQ
Q^{(k)}(\theta) =
G_{k,n+d}(\theta_{d}) G_{k,n+d-1}(\theta_{d-1})
\cdots G_{k,n+1}(\theta_{1}) \ .
\NEQ
An ORP from $O(d)$ to $e_d$ is
\BEQ \label{ORP3}
{\Qctl}_3 = \{ \Qtl(\theta)=  G_{1,2}(\theta_1) G_{2,3}(\theta_2) \cdots
        G_{d-2,d-1}(\theta_{d-2})       G_{d-1,d}(\theta_{d-1}) \
 \} \ ,
\NEQ
where now the angular restrictions are
 $-\pi /2 < \theta_i \le  \pi /2$ for $1 \le i < d-1$
and $-\pi  < \theta_{d-1} \le  \pi $.

\section{OTSON representations.} \label{OTSONRepSect}
The key to our OTSON representation is the recognition that the $(C,A)$
stack is column orthonormal. These results include the representation
of \cite{MRbook} as an important special case.
Our fundamental representation for OTSON pairs is

\begin{thm} \label{OTSONRepthm}
Every real OTSON pair has the representation:
\BEQ \label{CAdecomp}
\left( \begin{array}{c} C \\ A \end{array} \right) =
Q^{(n)}(\theta_n) Q^{(n-1)}(\theta_{n-1})  \ldots
Q^{(2)}(\theta_2) Q^{(1)}(\theta_1)
\left( \begin{array}{c} \Ib_n \\ 0_{d,n} \end{array} \right)
\NEQ
for some set of $n$ $d$-vectors $\{ \theta_1, \theta_2 \ldots \theta_n \}$.
Here the $Q^{(k)}$ are given by (\ref{QkStruct}) ,
where $ \mu_k$, $x_k$, $y_k$ and $\Otl_k$ are subblocks of
a ORP of $O(d+1)$ to $e_1$. 
\end{thm}

We  successively determine $\theta_n$, $\theta_{n-1}$, $\ldots$
$\theta_1$. At the $(n-k+1)$th stage, $\theta_k$  is determined
to zero out the $d$ of the $d+1$ nonzero entries
in the $k$th column.
By orthogonality the other entries in the $k$th row must be zero.

\Prf
We determine $\theta_n$ so that
$\left(Q^{(n)*}(\theta_n)Q \right)_{i,n}$$=\delta_{i,n}$.
By orthonormality, $\left(Q^{(n)*}(\theta_n)Q \right)_{n,i}$$=\delta_{i,n}$.
Let $\Omega^{(n+1)}$ be the $(C,A)$ stack and set
\BEQ
\Omega^{(k)}(\theta_{k}, \theta_{k+1}, \ldots ,\theta_{n}) \equiv
Q^{(k)*}(\theta_{k})  Q^{(k+1)*}(\theta_{k+1}) \ldots  Q^{(n)*}(\theta_n)
\left( \begin{array}{c} C \\ A \end{array} \right)  \ .
\NEQ
Assume that $\Omega^{(k)}$
has its last $(n-k+1)$ columns  satisfying $\Omega^{(k)}_{ij} =\delta_{ij}$.
Since  $\Omega^{(k)}$ has orthonormal columns,
$\Omega^{(k)}_{k:n, 1:(k-1)}= 0$. Select $\theta_{k-1}$ such
$\Omega^{(k-1)}_{(n+1):(n+d),(k-1)} =0$. Then
$\Omega^{(k-1)}_{j,(k-1)} =0$
for $j\ne k-1$ and therefore the
last $(n-k+1)$ columns  satisfy $\Omega^{(k-1)}_{ij} =\delta_{ij}$.
\eopp

For the Givens rotations of class ${\cal Q}_1$, we have
\BEQ
\left( \begin{array}{c} C \\ A \end{array} \right) =
G_{n,n+d}(\theta_{n,d}) G_{n,n+d-1}(\theta_{n,d-1})
\cdot G_{n,n+1}(\theta_{n,1})
\cdot \cdot
G_{1,n+d}(\theta_{1,d}) 
\cdot G_{1,n+1}(\theta_{1,1})
\left( \begin{array}{c} \Ib_n \\ 0_{d,n} \end{array} \right)
\NEQ

We now show that every matrix of the form given in the righthand side of
(\ref{CAdecomp}) is a OTSON matrix. We define
\BEQ
\Gamma^{(k)} \equiv
Q^{(k)}(\theta_{k})  Q^{(k-1)}(\theta_{k-1}) \ldots  Q^{(1)}(\theta_1)\ .
\NEQ

\begin{lem} \label{GamkThm}
Let $Q^{(k)}$ have the structure given
by (\ref{QtlStruct}) and (\ref{QkStruct}),
then $\Gamma^{(k)}$ has the structure:
\BEQ \label{GamkStruct}
\Gamma^{(k)} \equiv
\left(
\begin{array}{ccc}
L_k & 0_{k,n-k} & N_k \\
0_{n-k,k} & \Ib_{n-k} & 0_{n-k,d} \\
M_k & 0_{d,n-k} &P_k
\end{array} \right) \ ,
\NEQ
where $L_k$ is a lower triangular $k \times k$ matrix and
the following recurrence relations hold:
\begin{eqnarray}
L_{k} &=&
\left( \begin{array}{cc}
L_{k-1}& 0 \\
y_k^* M_{k-1} & \mu_k
\end{array}\right)
\ \ , \ \
N_{k} =
\left( \begin{array}{c}
N_{k-1} \\ y_k^* P_{k-1}
\end{array} \right)
\end{eqnarray}

\begin{eqnarray} \label{Minduct}
M_k = \left(\begin{array}{cc}
\tilde{O}_k M_{k-1} & x_k
\end{array} \right)
\ \ , \ \
P_k = \tilde{O}_k P_{k-1} \ .
\end{eqnarray}
\end{lem}

\Prf
Assume (\ref{GamkStruct}) for $k-1$ and multiply $Q^{(k)}\Gamma^{(k-1)}$.
\eopp

Lemma \ref{GamkThm} does not use the fact that $\Qtl^{(k)}$ is orthogonal.
Lemma \ref{GamkThm} is a special case of a more general theory
of matrix subblock products \cite{MR3}.

The last $(d+2)$ rows of $A$ may be rewritten as
\BEQ \label{Bot3}
A_{(n-d-1):n,1:n} =
\left(
\begin{array}{ccc}
y_{n-1}^* M_{n-2} & \mu_{n-1} & 0 \\
y_{n}^* \Otl_{n-1} M_{n-2} & y_{n}^* x_{n-1} & \mu_{n} \\
\Otl_{n}\Otl_{n-1}M_{n-2} & \Otl_{n} x_{n-1} & x_n
\end{array} \right) \ .
\NEQ
Lemma  \ref{GamkThm} implies that $L_n$ is lower triangular and
thus $\Gamma^{(n)}_{:, 1:n}$ corresponds to the $(C,A)$ stack
of an observer Hessenberg  system stack:

\begin{cor}
Every $(C,A)$ stack of the form (\ref{CAdecomp}) is a OTSON pair
when the $Q^{(k)}$ are orthogonal matrices satisfying (\ref{QkStruct}).
\end{cor}

A parameterization of state space models is identifiable
when only one parameter vector corresponds to each transfer function;
i.e.\ the map from parameters to input-output behavior is injective.
We now show that the mapping between standard OTSON pairs and
orthogonal product representation given in Theorem \ref{OTSONRepthm}
is one to one and onto.

\begin{thm} \label{HON1to1}
Let each $Q^{(k)}(\theta_k)$ be an embedding of a  
ORP of $O(d+1)$ to $e_1$ as given by (\ref{QtlStruct}).
Then there is a one to one correspondence between strict OTSON pairs and
the orthogonal product parameterization of Lemma \ref{GamkThm}
with $\{\mu_k > 0 \}$.
There is a one to one correspondence between unreduced OTSON pairs and
the orthogonal product parameterization restricted to $\{\mu_k \ne 0 \}$.
\end{thm}

\Prf
Theorem \ref{OTSONRepthm} shows that every OTSON pair has such
 a representation.
From (\ref{Minduct}), we represent the last $d$ rows of $A$ as
$M_n = \left(\Otl_n \Otl_{n-1} \cdot \cdot \Otl_2 x_1, \ldots \Otl_n x_{n-1}, x_n \right)$.
For $\mu_k \ne 0$, $\Otl_k$ may be determined and inverted and $x_{k-1}$
is determinable from $A_{k-1:n,n-d+1:n}$.
We let $\{ x_1, x_2 \ldots x_n \}$ vary over $|x_k|<1 $.
Thus the mapping of OTSON pairs into the product
ORP representation is {\em onto}.
\eopp


For our parameterization of output pairs to be truly identifiable, we need to
restrict our parameter space, $\theta \in \Theta$, such that
no two output pair representations, $Q(\theta_1)$ and $Q(\theta_2)$, are
equivalent. We prefer to restrict our parameterizations to
 $\{\theta |\mu_k \ge 0\}$. This set has redundant representations only
when at least one $\mu_k =0$.



\section{HESSENBERG OBSERVER OUTPUT NORMAL FORM} \label{HOONRepSect}

In this section, we give representation results for HOON pairs.
The first row of $C$ satisfies $C_{1,1} = \sqrt{1 - \gamma^2}$,
$C_{1,j} = 0$ for $j>1$. We do not transform this row and treat $\gamma$ as a free parameter.
We use Givens rotations to zero out  the lower diagonal of $A$ and the row $2$ through
row $d$ of $C$. For each column of the $(C, A)$ stack, we use $d$ Givens rotaions
except for the final column wich requires only $d-1$.

We embed orthogonal reduction parameterizations of $O(d+1)$ to $e_{d+1}$ into the
space of $(n+d -1) \times (n+d -1) $ matrices.
We define $\Vtl(\theta)$ in $(n+d -1) \times (n+d -1) $ dimensional matrices
$V^{(k)}$:
\BEQ \label{VkStruct}
V^{(k)}(\theta) =
\left(
\begin{array}{ccc}
\tilde{O}_{k} &0_{d, k-1} &  x_k \\
 0_{k-1,d} & \Ib_{k-1} & 0_{k-1,1} \\
 y_k^* & 0_{1,k-1} & \mu_k
\end{array}\right)
\oplus \Ib_{n-k-1} \ ,
\NEQ
for $1 \le k <n$. Here $x_k$, $y_k$ are $d$-vectors.
Thus $V^{(k)}$ alters only the rows $1:d$ and row $k+d$.
We require that the $(d+1) \times (d+1)$ orthogonal matrix,
\BEQ \label{VtlStruct}
\Vtl(\theta) \equiv
\left( \begin{array}{cc}
\Otl_k& x_k \\
y^*_k & \mu_k
\end{array}\right) \
\NEQ
be a member of a ORP from $O(d+1)$ to $e_{d+1}$.
For $V^{(n)}(\theta)$, we define.
\BEQ \label{VnStruct}
V^{(n)}(\theta) =
\Vtl_d(\theta) \oplus \Ib_{n-1} \ ,
\NEQ
where $\Vtl_d(\theta)$ is a ORP from $O(d)$ to $e_d$.
Thus $\{ \theta_1, \ldots \theta_{n-1}\}$ are $d$-vectors while $\theta_n$ is
a $d-1$-vector. Our parameterization of HOON pairs uses a scalar, $0 \le \gamma<1$
and $\{ \theta_1, \ldots \theta_{n}\}$.
We denote the bottom $(d-1)$ rows of $C$ by $\Ch$.

\begin{thm} \label{HOONRepthm}
Every real nondegenerate HOON pair 
has the representation:
\BEQ \label{HOONRepEq}
\left( \begin{array}{c} \Ch \\ A \end{array} \right) =
V^{(1)}(\theta_1) V^{(2)}(\theta_{2})  \ldots
V^{(n-1)}(\theta_{n-1}) V^{(n)}(\theta_n)
\left( \begin{array}{c} 0_{d-1,n} \\ P(\gamma) \end{array} \right)
\NEQ
for some set of parameters, $\{\gamma, \theta_1, \theta_2 \ldots \theta_n \}$,
with $0<|\gamma| <1$ and $C_{1,1} = \sqrt{1 - \gamma^2}$.
Here $P(\gamma)$ is the $n \times n$ scaled permutation matrix:
$P_{2,1}= \gamma$, $P_{k+1,k}=1 $ for $2 \le k <n$, $P_{1,n}=1$,
and $P_{i,j}=0$ otherwise. The $V^{(k)}(\theta_k)$ are defined in
(\ref{VkStruct})-(\ref{VnStruct}) 
and are members of the appropriate ORPs.
\end{thm}


\Prf
Let $\Omega_{n+1}$ be the $(\Ch,A)$ stack and set
\BEQ
\Omega^{(k)}(\theta_{1}, \theta_{2}, \ldots ,\theta_{k}) \equiv
V^{(k)*}(\theta_{k})  V^{(k-1)*}(\theta_{k-1}) \ldots  V^{(1)*}(\theta_1)
\left( \begin{array}{c} \Ch \\ A \end{array} \right)  \ .
\NEQ
Assume that $\Omega^{(k)}$
has its first $k$ columns  satisfying
$\Omega^{(k)}_{i,j} =\gamma_j \delta_{i-d,j}$,
where $\gamma_1=\gamma$ and $\gamma_j=1$ for $1<j\le k$.
Since  $\Omega^{(k)}$ has orthonormal columns,
$\Omega^{(k)}_{(d+1):(d+k), 1:n}= \gamma_j \delta_{i-d,j}$.
Select $\theta_{k+1}$ such
$\Omega^{(k+1)}_{1:d,k+1} =0$. Then
$\Omega^{(k+1)}_{j,(k+1)} =0$
for $j\ne k+1$, and therefore the
first $k+1$ columns  satisfying $\Omega^{(k+1)}_{ij} =\delta_{ij}$.
\eopp

For $d=1$ and $C_{1,:}=0$, (\ref{HOONRepEq}) is the well-known expression of an
unitary Hessenberg
matrix as a product of $n$ Givens rotations \cite{AGR}.
To show that every matrix of the form given by the righthand side of
(\ref{HOONRepEq}) is a HOON pair, we define

\BEQ \label{Xdef}
X^{(k)}(\theta_1,\theta_2, \ldots , \theta_k)  \equiv
V^{(1)}(\theta_1)V^{(2)}(\theta_{2})  \ldots
V^{(k)}(\theta_k) \ .
\NEQ

\begin{lem} \label{XkThma}
Let $V^{(k)}$ have the structure given
by (\ref{VkStruct}) - (\ref{VnStruct}),
then $X^{(k)}$ has the structure:
\BEQ \label{XkStruct}
X^{(k)} \equiv
\left(
\begin{array}{cc}
N_k  & H_k
\end{array} \right)
\oplus \Ib_{n-k-1} \ ,
\NEQ
where $N_k$ is $(d+k)\times d$ and
$H_k$ is a $(d+k)\times k$ upper triangular 
matrix and the following recurrence relations hold:
\begin{eqnarray}
N_{k} =
\left( \begin{array}{c}
N_{k-1} \Otl_{k} \\ y_k^*
\end{array} \right)
\ \ , \ \
H_{k} &=&
\left( \begin{array}{cc}
H_{k-1}& N_{k-1} x_k \\
0_{1,k-1} & \mu_k
\end{array}\right) \ .
\end{eqnarray}
\end{lem}

This result follows from multiplying out the matrix product.

\begin{cor} \label{XkThm}
Let $V^{(k)}$ have the structure given
by (\ref{VkStruct}) - (\ref{VnStruct}) and let $P(\gamma)$
be the scaled permutation matrix. The righthand side of
(\ref{HOONRepEq}) defines an HOON pair with
$C_{1, j} = \sqrt{1-\gamma^2} \delta_{1,j}$ and $C_{2:d,:}= \Ch$.
\end{cor}


\begin{thm}
Under the definitions of  Theorem \ref{HOONRepthm},
there is a one to one correspondence between strict HOON pairs and
the parameterization of Theorem \ref{HOONRepthm}
restricted to $\{\mu_k > 0 \}$.
\end{thm}

\Prf
From Lemma \ref{XkThma},
$A_{2,1} = \gamma \mu_1$ and $A_{k+1,k} = \mu_k$ for $2\le k <n$.
The first $d$ rows of $A$ as
$H^{(n)}_{1:d,1:n} =
\left(x_1, \Otl_1 x_2, \Otl_1 \Otl_2 x_3, \ldots
 \Otl_1\cdot \Otl_{n-1} x_n \right)$.
Thus we can determine $x_{k+1}$ from $A$ and $\{x_1,\ldots x_k\}$.
The proof is now identical to the proof of Theorem \ref{HON1to1}.
\eopp

\section{DISCUSSION}

For each of the three output pairs, Schur ON form,
observer triangular system ON
and  Hessenberg observer ON pairs,
we have examined the uniqueness/identifiability of the representation in
Section \ref{UniqSect}. 
We then express each of these output pairs in terms of
an orthogonal product representation (OPR)
as the product of orthogonal matrices involving a total of $nd$ parameters
(Theorems \ref{OTSONRepthm}, \ref{HOONRepthm}). 
A similar representation is possible for Schur ON form \cite{MR5}.
We have shown how to place restrictions on the parameters such that
the orthogonal product representations are in one to one correspondence
with sets of generic transfer functions.
For OTSON and HOON representations, we recommend restricting the
Given's rotations in  Theorems \ref{OTSONRepthm} and \ref{HOONRepthm}
such that
 $\{\theta |\mu_k \ge 0\}$. This set has redundant representations only
when at least one $\mu_k =0$.

In practice, these orthogonal product representations
are implemented with either Given's rotations or Householder transformations.
Our definition of ORPs allows us to treat all the standard cases
similarly.
We do not explicitly store or multiply by $Q^{(k)}$ or $V^{(k)}$.
Instead we store only the Given's or Householder parameters and
we perform the matrix multiplication implicitly. For an $n$-vector $v$,
we compute $Av$ and $Cv$ using the  orthogonal product representation.

These orthogonal product representations have several advantageous properties:

1) $ \frac{d}{d \theta_k} \left( \begin{array}{c} C \\ A \end{array} \right)$
is easy to compute.


2) Vector multiplication by
$Q$ 
and by $ \frac{d}{d \theta_k} Q$ require $O(6nd)$ and $O(8nd)$ operations,
where $Q$ is the $(C,A)$ stack.

3) Observability and stability are equivalent and
$\|A \| \le 1$ is automatically satisfied

4) The controllability matrix, $B$, may be parameterized by its elements, $B_{i,j}$,
separately from the parameters of $(A,C)$.

5) The observability Grammian is perfectly conditioned.

The final advantage is key for us.
Many of the other well-known representation are very ill-conditioned
\cite{MR4}.
A measure of the conditioning of a representation is the product
of the condition number of the observability Grammian and the
condition number of the controllability Grammian.
As discussed in \cite{MR4}, balanced, input normal and output normal
representations minimize this product of the condition numbers.

The fast filtering methods of \cite{VDV} may be further sped up when
 $(A,C)$ or $(A^*,B)$ has
the orthogonal product representations of this article. To transform
a specific output pair to ON form, the dual Stein equation must be solved.
The numerical conditioning of this problem can be quite poor
\cite{PenzlBnd,MR4}.

Which orthogonal product representation is most appropriate for my problem?
Schur ON representations naturally display the eigenvalues of $A$ while
the spectrum of $A$ must be numerically calculated when $A$ is OTSON or
HOON. If the parameterization evolves in time, the form of the
Schur representation changes when eigenvalues coalesce and the block structure
of $A$ changes. Thus, for evolving representations, we prefer the
OTSON and HOON representations. It is straightforward to impose
the restrictions that ${\rm rank\ }C=d$ in the OTSON form.
If the problem requires derivatives of $A(\theta)$ and $C(\theta)$,
the Givens rotation parameterization of ORPs is usually simpler than
Householder reflections.

In summary, these orthogonal product representations offer the best possible
conditioning while having a convenient representation with fast
matrix multiplication.
Corresponding controller representations
exist for input pairs, $(A,B)$, that are input normalized.


\section{APPENDIX A: SPECIFYING THE REAL SCHUR FORM}

The real Schur representation is defined and described
in \cite{Demmel,Edelman,GVL}.
We denote the number of complex conjugate
pairs of eigenvalues by $\ell$ and the number of real eigenvalues by $n-2\ell$.
Let
$m(k) = 2k-1$ for $k \le \ell$ and
 $m(k) = k + 2\ell $ for $\ell < k \le n -\ell + 1$ with $m(0)=0$,
and define $M=n -\ell$. 
The Schur form is
\BEQ \label{SchurDef}
\left(
\begin{array}{ccccc}
Z_1& R_{1,2} & R_{1,3} & \ldots & R_{1,M} \\
0 & Z_2 &  R_{2,3}&  \ddots & R_{2,M} \\
 0 & \ldots & \ddots &  \ddots & \vdots\\
\vdots  &  \ldots &0& Z_{M-1} & R_{M-1,M} \\
0 & \ldots      & 0 & 0& Z_M \\
\end{array}
\right) \ ,
\NEQ
where $Z_i$ are $2 \times 2$ matrices for $i \le \ell$ and
real scalars for $\ell< i \le n- \ell$. Here
 $R_{i,j} \equiv A_{m(i):(m(i+1)-1),m(j):(m(j+1)-1)}$.
Thus we explicitly require the complex conjugate eigenvalues to be
placed ahead of the real eigenvalues for a matrix to be in Schur form.
For identifiability, we need to  uniquely specify the order of the
blocks and the form of each block. Let $\{\lambda_i \}$ be the eigenvalues
of $A$ with $\lambda_{m(k)}$ being an eigenvalue of $Z_k$ and
 $\lambda_{2j}$ being an eigenvalue of $Z_j$ for $j \le \ell$.

\begin{defn}\label{orderSchur}
Let $A$ be in real Schur form, (\ref{SchurDef}), with ordered eigenvalues
$\{\lambda_j\}$ as described above.
Then $A$
is in ordered Schur form if 1) $|\lambda_j| \ge |\lambda_i|$ for
 $i <j \le \ell$ and for $2\ell <i <j$;
2) If  $|\lambda_m(j)| = |\lambda_m(i)|$, then
 ${\rm Re\ } \lambda_m(j) \le {\rm Re\ } \lambda_m(i)$
for  $i <j \le \ell$ and for $2\ell <i <j$;
\end{defn}

Definition \ref{orderSchur} can be replaced by
any other complete specification of the eigenvalue block order.
Note that  $A$ may be transformed by a product of $\ell$ Givens rotations:
 $G_{1,2} G_{3,4} \cdots G_{2\ell-1, 2\ell}$ and still stay in Schur form.
For identifiability,
we also need to specify the form of each $2\times 2 $ diagonal subblock.
Let $Z_j$ denote a $2\times 2$ diagonal subblock of a Schur $A$:
\BEQ
Z_j =
\left( \begin{array}{cc} z_{11} & z_{12} \\ z_{21} & z_{22}
\end{array}\right) \ .
\NEQ
A common standardization of $Z_j$ is to require $z_{11} =z_{22}$,
with $z_{12}z_{21}<0$ and $z_{12} + z_{21}>0$.
We refer to this standardization of the two by two
subblocks as $\lambda_r$ block form  since $z_{11} =z_{22}= \lambda_r$,
the real part of the eigenvalues.
The $\lambda_r$ form is also known as standardized form \cite{BaiDemmel}.

\begin{thm} \label{SchurUniq}
Let $A$ and $\Ah$ be $n\times n$ matrices in real Schur form
with ordered eigenvalues. Let $A$ and $\Ah$ be orthogonally
similar: $\Ah U =U{A}$ with $U$ orthogonal.
Let $m$ be the number of distinct eigenvalue pairs
plus the number of distinct real eigenvalues. Partition
$A$, $\Ah$ and $U$ into $m$ blocks corresponding to the
repeated eigenvalue blocks. Then $U$ has block diagonal form:
 $U = U_1 \oplus U_2 \oplus \ldots \oplus U_m$, where $U_i$ is orthogonal.
\end{thm}

\Prf
From
$ \hat{A}_{m, m} U_{m,1} =  U_{m,1} A_{1,1}$.
If $m>1$, then $ \hat{A}_{m, m}$ and $ A_{1,1}$ have no
common eigenvalues. By Lemma 7.1.5 of \cite{GVL}, $U_{m,1}\equiv 0$.
Repeating this argument shows $U_{m-k,1}\equiv 0$
 for $k=0,1\ldots <m-1$.
By orthogonality, $U_{1,j}= 0$ for $1<j<m$.
We continue this chain
showing that  $U_{i,2} = 0$ for $i \ne 2$, etc.
Proof by finite induction.
\eopp

When $A$ has the distinct eigenvalues the block decomposition
if precisely that of (\ref{SchurDef}). When $A$ has
eigenvalue with multiplicity greater than one, the block decomposition groups
the repeated eigenvalue blocks together.
In the repeated eigenvalue case, not every block orthogonal transformation,
 $U = U_1 \oplus U_2 \oplus \ldots \oplus U_m$, preserves the
Schur form. We use the freedom of the $2 \times 2$ orthogonal blocks
to standardize the diagonal of the real Schur form:

\begin{cor} \label{SchurUniqCor}
Let $A$ be a $n\times n$ matrix with distinct eigenvalues.
The $A$ is orthogonally similar to a matrix, $\hat{A}$,
in ordered $\lambda_r$ real Schur form.
and $\hat{A}$ is unique up to diagonal
unitary similarities:
 $\hat{A} \leftarrow E {A}E^*$,
where $|E_{i,j}|^2=\delta_{i-j}$ and $E_{j,j}=1 $ for $j\le 2\ell$.
\end{cor}


\Prf
Existence of the orthogonal transformation is proven in
Theorem 2.3.4 of \cite{HJ}
and by Theorem \ref{SchurUniq}, it is unique up to block orthogonal transformations.
The $\lambda_r$ standardization
uniquely determines the $2 \times 2 $ diagonal subblocks of $A$ \cite{Edelman}.
\eopp


We propose the alternative standardization of the $2\times 2$ blocks:

\begin{defn}
Let $A$ be a Schur matrix as given by (\ref{SchurDef}).
It is in {\em qd} block form if each $2\times 2$ diagonal subblock
satisfies  $Z_j = Q_j D_j$, where  $Q_j$ is a $2\times 2$ orthogonal matrix
and $ D_j$ is a nonnegative diagonal matrix:
\BEQ \label{qdEq}
Z_j =
\left( \begin{array}{cc} c & s \\ -s & c
\end{array}\right)
\left( \begin{array}{cc} d_{1} & 0 \\ 0 & d_{2}
\end{array}\right) \ ,
\NEQ
with $s \ge 0$ and $d_1 \ge d_2$.
\end{defn}

The {\em qd } condition implies as
$z_{11}  z_{12} + z_{21}  z_{22}=0$, $z_{12} \ge 0$.
 An uniqueness result is

\begin{lem} \label{QDlem}
Every $2 \times 2$ nonsingular matrix, $M$, is orthogonally similar
to an unique matrix $Z=QD$ in {\em qd} block form.
\end{lem}

\Prf
Let $M$ have the singular value decomposition, $M = U\Lambda V^*$ and set
$Q = V^*U$ and $D=\Lambda$.
If $d_1<d_2$, then permute the rows and columns of $Z$:
 $D \rightarrow PDP$,  $Q \rightarrow PQP$, where $P$ is
the $2\times 2$ permutation matrix: $P_{1,2}=P_{2,1}=1$
and $P_{1,2}=P_{2,1}=0$.
If $s<0$, orthogonally transform $Z$ by $T={\rm diag\ } (1,-1)$.
Now suppose $Z_1$ and $Z_2$ are both in  {\em qd} block form
and are both orthogonally similar to $M$. Then
Let $Z_i =Q_i D_i$ for $i=1,2$. Then there are two orthogonal matrices
$U_1$ and $U_2$ such that
$D_1 = U_1^* D_2 U_2$.
From $D_1^2 = U_2^* D_2^2 U_2$, $D_1=D_2$.
If $D_{i; 1,1} \ne  D_{i;2,2}$ for $i=1,2$, then $U_1=U_2=\Ib_2$.
If $D_{i; 1,1} = D_{i;2,2}$, then by direct computation, $Q_1=Q_2$.
\eopp

Since we can always rotate the $2\times 2 $ diagonal subblocks
from $\lambda_r$ form to $qd$ form, we have

\begin{cor}
Let $A$ be a stable $n\times n$ matrix with distinct eigenvalues.
The $A$ is orthogonally similar to an unique matrix $\hat{A}$,
where  $\hat{A}$ is ordered $qd$ real Schur form.
\end{cor}

\bibliographystyle{IEEE}
\end{document}